\documentclass[prd,twocolumn,a4paper,altaffilletter]{revtex4}
\usepackage{graphicx}% Include figure files
\usepackage{bm}% bold math

\newcommand{\be}{\begin{equation}}
\newcommand{\ee}{\end{equation}}
\newcommand{\ba}{\begin{eqnarray}}
\newcommand{\ea}{\end{eqnarray}}

\begin{document} 

%\title{Linear momentum in quasistatic electromagnetic systems}
\title{Tracking solutions in tachyon cosmology}
\author{J.M. Aguirregabiria} \email{wtpagagj@lg.ehu.es}
\author{R. Lazkoz} 
\email{wtplasar@lg.ehu.es}
\affiliation{Fisika Teorikoa, Zientzia eta Teknologiaren Fakultatea, Euskal Herriko Unibertsitatea,
644 Posta Kutxatila, 48080 Bilbao, Spain} 

%%%%%%%%%%%%%%%%%%%%%%%%%%%%%%%%%%%%%%%%%%%%%%
\begin{abstract} 
We perform a thorough phase-plane analysis of the flow defined by the equations of motion 
of a FRW universe filled with a tachyonic fluid plus a barotropic one. The tachyon potential is assumed to be of inverse square form, thus allowing for a two-dimensional autonomous system of equations. The Friedmann constraint, combined with a convenient choice of coordinates, renders the physical state compact. We find the fixed-point solutions, and discuss whether they represent attractors or not. The way the two fluids
contribute at late-times to the fractional energy density depends on how fast the barotropic fluid redshifts. If it does it fast enough, the tachyonic fluid takes over at late times, but if the opposite happens,  the situation will not be completely dominated by the barotropic fluid; instead there will be a residual non-negligible contribution from the tachyon   subject to restrictions coming from nucleosynthesis.
\end{abstract} 

\maketitle

%\pacs{PACS: 45.20.Jj, 01.55.+b}

\section{Introduction}
A phase of accelerated inflation in the early stages of our universe is favored by
first-year WMAP data \cite{WMAP}. Thus, it seems  inflation is here to stay as the dominant paradigm for structure formation. The quest for a string theory motivated explanation of cosmological inflation has resulted
in the emergence of the proposal of inflation driven by a tachyon field. The idea strongly relies in the possibility of describing tachyon condensates
in terms of perfect fluids within string theories \cite{sen}. A plethora of papers studying cosmological consequences of such  fluids have appeared since, some in the framework of general relativity  \cite{power-law,others, exponential, inverse}, some others in the brane-world scenario \cite{brane}. 

As happens with standard scalar fields, one's favorite inflationary behaviour is tailored
by {\it ad hoc} choices of the initial conditions and the shape of the potential, but it is important to investigate up to what extent the features of the model depend on those choices. One way to address that problem is to consider tachyon field dynamics, because  for a given potential such an analysis will provide us with constraints on the initial conditions.

The stability of tachyonic inflation  against changes in initial conditions  has been studied  for an exponential potential  \cite{exponential} and for the inverse power-law potential \cite{inverse}. Exact solutions for a purely tachyonic matter content with an inverse square potential
are known \cite{power-law}, but no solutions exist for cases which combine tachyonic and barotropic fluids, so a dynamical systems approach may be relevant. Interestingly, the inverse square potential plays the same role for tachyon fields as the exponential potential \footnote{The general solution to the Einstein equations for a FRW spacetime with an exponential potential  was given in \cite{solution}.} does for standard scalar fields \cite{liddle,firstsem,exp,thirdsem}. On the one hand, those are the potentials that give power-law solutions.  On the other hand, only those potentials allow constructing a two-dimensional autonomous system  \cite{thirdsem} using the evolution equations, whereas for any other potential the number of dimensions will be higher if the system is to remain autonomous.

As compared to \cite{inverse},  we throw in one more ingredient in our study, because  we allow for the presence of a barotropic fluid, along with the tachyon fluid. 
The crucial consequence is the appearance of fixed-point solutions in which  the two fluids redshift at the same rate (tracking behaviour \cite{liddle}), so that there is some sort of equilibrium.  Tracking solutions are particularly interesting because their dynamical effects mimic a decaying cosmological constant (see \cite{thirdsem,firstsem,secondsem} for seminal references). Now, the fine-tuning problems posed by a cosmological constant would be waived precisely because of the independence on the initial conditions. Nevertheless, the contribution
of such relics to the fractional energy density are bounded by nucleosynthesis. 

On top of those interesting features, tracking solutions act as attractors at large, which means that the system is not any picky in what initial conditions are regarded. 

In Section II we study the phase-plane, find its fixed points and characterize them. In Section III we discuss the cosmological consequences of the attractor solutions: in subsection III A we consider tachyon dominated  solutions, whereas in subsection III B we
discuss the tracking ones. Finally, in Section IV we outline our main conclusions and future prospects.

%%%%%%%%%%%%%%%%%%%%%%%%%%%%%%%%%%%%%%%%%%%%%%
\section{Phase plane}\label{sec:equations}
%%%%%%%%%%%%%%%%%%%%%%%%%%%%%%%%%%%%%%%%%%%%%%

The evolution equations for a flat ($k=0$) Friedmann-Robertson-Walker (FRW) cosmological
model filled with a tachyon field $T$ evolving in a potential $V(T)$ and a barotropic perfect fluid with equation of state $p_{\gamma}=(\gamma-1)\rho_{\gamma}$ are
\begin{equation}
-2\dot H=\frac{V\,\dot T^2 }{\sqrt{1-\dot T^2}}+\gamma\rho_{\gamma},\label{eq:raychaudhuri}
\end{equation}
\begin{equation}
\frac{\ddot T}{1-\dot T^2}+3H\dot T+\frac{V,_T}V=0,\label{eq:kleingordon}
\end{equation}
\begin{equation}
\dot\rho_{\gamma}+3\gamma H\rho_{\gamma}=0,\label{eq:cont}
\end{equation}
which are, in turn, subject to the Friedmann constraint
\be
3 H^2=\frac{V}{\sqrt{1-\dot T^2}}+\rho_{\gamma}\label{friedmann}.
\ee
Here and throughout overdots denote differentiation with respect to cosmic time $t$, $H\equiv  \dot a/a $ is the 
Hubble parameter, and $a$ is the synchronous scale factor. 

One can also define an energy density $\rho_T$ and a pressure $p_T$ for the tachyon, so that it can be thought of as a perfect fluid. We have then
\begin{eqnarray}
\rho_T=\frac{V}{\sqrt{1-\dot T^2}},\\
p_T=-V\sqrt{1-\dot T^2}.
\end{eqnarray}

If we use $\log a^3$ as independent variable instead of the cosmological time, for any time dependent function $f$ we get
 \begin{equation}
 f'=\frac{\dot f}{3H}.
 \end{equation}
As usual, we also introduce n convenient variables:
 \begin{eqnarray}
  x&\equiv&\dot T,\\
  y&\equiv&\frac{V}{3H^2},\\
  z&\equiv&\frac{\rho_{\gamma}}{3H^2}.
 \end{eqnarray}

\begin{figure}
\begin{center}
\includegraphics[width=0.4\textwidth]{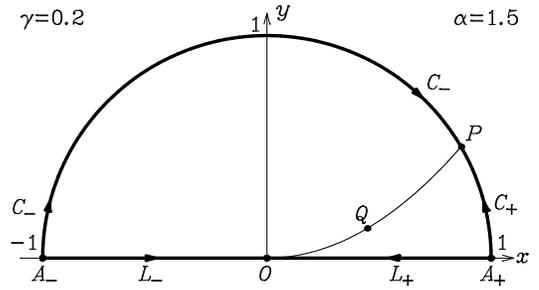}
\end{center}
\caption{Toroidal solenoid and point charge.\label{fig:phase}} 
\end{figure}

Let us concentrate now on the inverse square potential $V=\beta T^{-2}$. The evolution of the model is described by the dynamical system
 \begin{eqnarray}
 x'&=&\left(x^2-1\right)\left(x-\sqrt{\alpha y}\right),\\
 y'&=&y\left[x \left(x-\sqrt{\alpha y}\right)+z\left(\gamma-x^2\right)\right],\\
 z'&=&z(z-1)\left(\gamma-x^2\right),
 \end{eqnarray}
along with the constraint
 \begin{equation}
  \frac{y}{\sqrt{1-x^2}}+z=1,\label{eq:constraint}
 \end{equation}
 which renders the phase space two dimensional so that we may speak of phase plane.
 For the sake of simplicity, we are using the following definition:
 \begin{equation}
 \alpha\equiv\frac 4{3\beta}>0. 
 \end{equation}
The physical constraints $V>0$ and $\rho_{\gamma}>0$ set limits on the dependent variables:
$-1<x<1$, $y>0$ and $x^2+y^2\le1$. In consequence, the phase plane is  restricted to the upper half of the unit disk
centered at the origin, as depicted in Fig.~\ref{fig:phase}. For convenience we will also include the whole
segment $-1\le x\le 1$, $y=0$ in our phase plane.

In addition, one can define a barotropic index for the tachyon fluid:
\begin{equation}
\gamma_T\equiv\frac{\rho_{T}+p_T}{\rho_T}.
\end{equation}
and, provided $V\ne 0$, one gets $\gamma_T=\dot T^2$.

Depending on the values of $\gamma$ and $\beta$ there may be up to
five fixed points ($O$, $A_\pm$, $P$ and $Q$) and up to 
six
heteroclinic orbits that connect pairs of fixed points ($L_\pm$, $C_\pm$, $M_\pm$). 
The fixed point $O$ located at the origin $(x,y)=(0,0)$ corresponds to $z=1$ and is a unstable
saddle (except in the very particular case in which $\gamma=0$, which will be discussed below).
The orbits  $L_\pm$ in the stable manifold correspond to the characteristic exponent $\lambda_1=-1$
while the exponent for the unstable space is $\lambda_2=\gamma/2$.

The fractional densities of the two fluids are respectively defined as :
\begin{eqnarray}
&&\Omega_{\gamma}\equiv\frac{\rho_{\gamma}}{3H^2}=z,\\
&&\Omega_{T}\equiv\frac{\rho_{T}}{3H^2}=\frac{y}{\sqrt{1-x^2}}.
\end{eqnarray}\begin{figure}
\begin{center}
\includegraphics[width=0.4\textwidth]{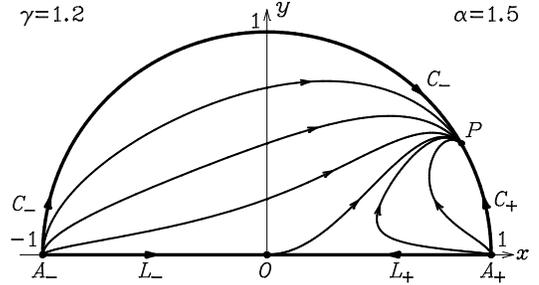}
\end{center}
\caption{Phase-space for $\alpha=1.5$ and $\gamma=1.2$.\label{fig:g12a15}} 
\end{figure}

\begin{figure}
\begin{center}
\includegraphics[width=0.4\textwidth]{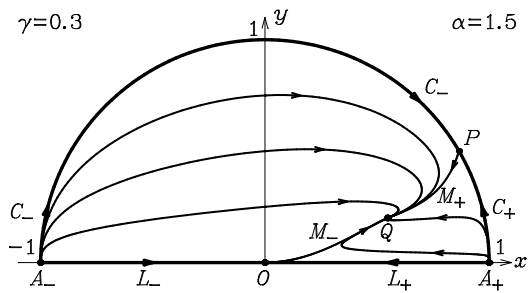}
\end{center}
\caption{Phase-space for $\alpha=1.5$ and $\gamma=0.3$.\label{fig:g03a15}} 
\end{figure}

The fixed points $A_\pm$ located at $(x,y)=(\pm1,0)$ correspond to $z=0$, and are unstable
nodes.  The orbits $L_\pm$ in the unstable manifold correspond to the characteristic
exponent $\lambda_1=2$ and orbits $C_\pm$ to $\lambda_2=\gamma/2$.

The fixed point $P$ at $(x,y)=(\sqrt{\alpha y_1},y_1)$, with
 \begin{equation}
 y_1\equiv\frac{\sqrt{\alpha^2+4}-\alpha}2,
 \end{equation}
always admits the arcs $C_\pm$ as the orbits in its stable manifold corresponding to the
 characteristic exponent $\lambda_1=-1+\alpha y_1/2<0$, 
 while the second exponent is $\lambda_2=\alpha y_1-\gamma$.
In consequence, $P$ is an asymptotically stable node for $\gamma>\gamma_1\equiv\alpha y_1$, in which case 
the phase space looks as depicted in Fig.~\ref{fig:g12a15}: $P$ is a global attractor, nearly all solutions
ends there.
%%%%%%%%%%%%%%%%%%%%%%%%%%%%%%%%%%%%%%%%%%%%%%
When $\gamma=\gamma_1$ a bifurcation arises: $P$ turns into a unstable saddle and at the same time
there appears a new attractor $Q$, which moves from $P$ to $O$ along an arc of the parabola 
$(x,y)=(\sqrt\gamma,\gamma/\alpha)$ as $\gamma$ decreases from $\gamma_1$ to $0$
(see Fig.~\ref{fig:phase}). The characteristic exponents
are 
 \begin{equation}
\lambda=\frac{\alpha\left(\gamma-2\right) \pm 
    \sqrt{16\alpha\gamma^2\sqrt{1-\gamma}+\alpha^2\left(4-20\gamma+17\gamma^2\right) }}{4\alpha}, 
 \end{equation}
so that $\Re\alpha <0$ for all $0\le\gamma<\gamma_1$, $Q$ is always asymptotically stable and  is a node (focus)
when the argument of the square root is positive (negative). A particular case is shown in
Fig.~\ref{fig:g03a15}.

\begin{figure}[b!]
\begin{center}
\includegraphics[width=0.4\textwidth]{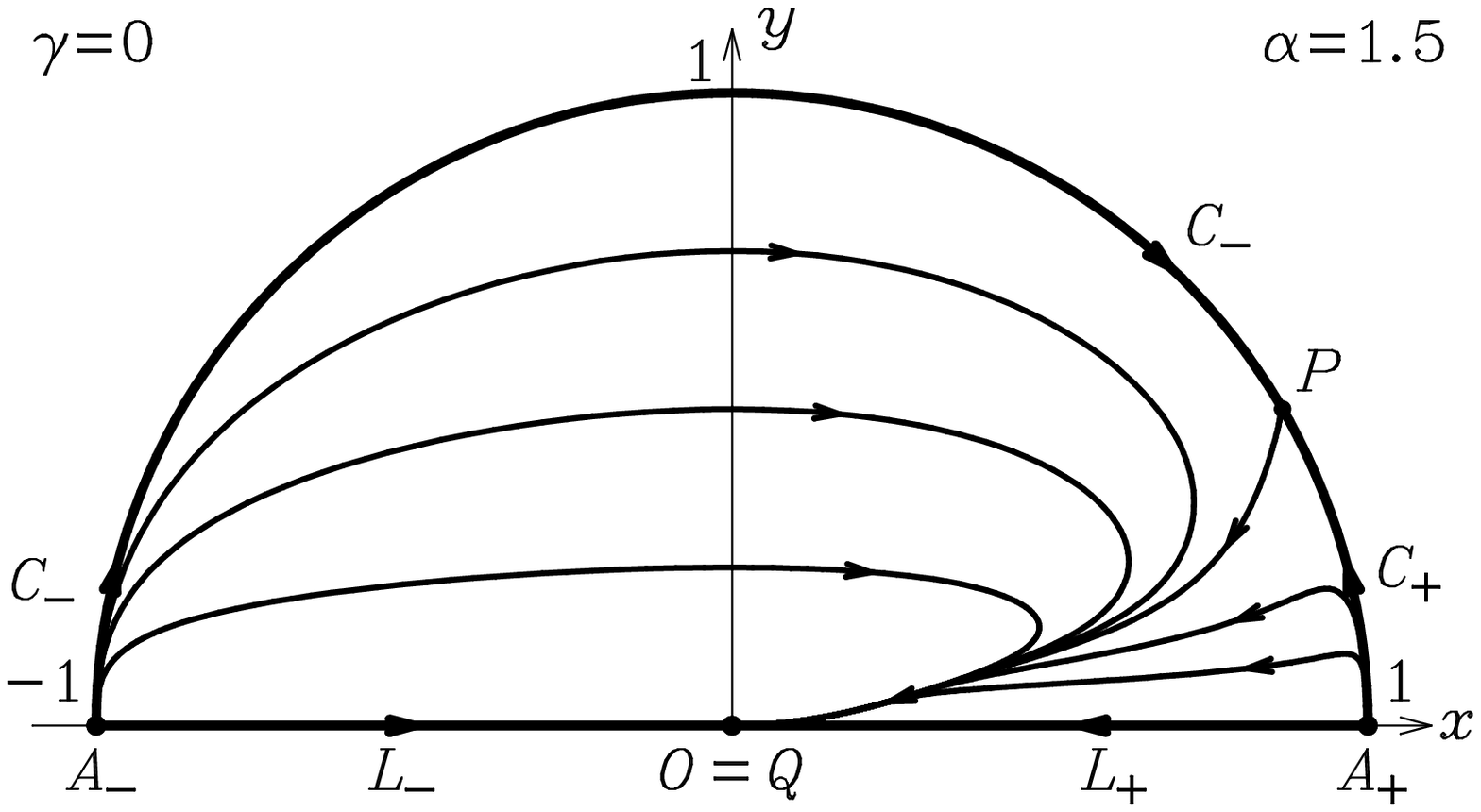}
\end{center}
\caption{Phase-space for $\alpha=1.5$ and $\gamma=0$.\label{fig:g00a15}} 
\end{figure}
%\newpage

In the limit case in which $\gamma=0$, fixed points $Q$ and $O$ coincide and are the attractor
in the system, as depicted in Fig.~\ref{fig:g00a15}.

\begin{widetext}
\begin{center}
\begin{table}[h!]\caption[crit]{\label{crit} The properties of the critical points.}
\begin{tabular}{ccccccc}
\hline
\hline
Name &$x$&$y$&\quad Existence\quad& Stability&$\Omega_T$&$\gamma_T$\\
\hline
$O$&0&0&\quad All $\gamma$ and $\beta$\quad & \quad  Unstable saddle for  $\gamma\ne0$\quad &0& Undefined\\
&&&&Stable node for $\gamma=0$&\\
$A_+$&$1$&$0$&\quad All $\gamma$ and $\beta$\quad &\quad Unstable node\quad &1&1\\
$A_-$&$-1$&$0$&\quad All $\gamma$ and $\beta$\quad&\quad Unstable node\quad &1&1\\
$P$&$\sqrt{\alpha y_1}$&$y_1$&\quad All $\gamma$ and $\beta$\quad&\quad Stable node for $\gamma<\gamma_1$\quad &1&$\alpha y_1$\\
&&&&\quad Unstable saddle for $\gamma\ge\gamma_1$\quad \\
$Q$&$\sqrt{\gamma}$&$\displaystyle\frac{\gamma}{\alpha}$&$\gamma<\gamma_1$&\quad Stable node\quad &$\displaystyle\frac{\gamma }{\alpha \,{\sqrt{1 - \gamma }}}$&$\gamma$\\
\vspace{-0.35cm}\\
\hline
\hline
\end{tabular}
\end{table}\end{center}
\end{widetext}

\section{Cosmological features of the attractor solutions }
Now we turn our attention to  the effective equation for the tachyon fluid in
the attractor solutions, so that we can discuss in broad terms what sort of evolution
they give rise to.
\subsection{Tachyon dominated solutions}
The attractor solutions $(x,y)=(\sqrt{\alpha  y_1},y_1)$ depict a situation in which the
energy density of the fluid vanishes, so it will be referred to as the tachyon dominated solution.
It is straightforward to see it corresponds to
\begin{equation}
\gamma_T=\alpha y_1
\end{equation}
and
\begin{equation}
T=\sqrt{\alpha y_1}\, t+T_0,
\end{equation}with $T_0$ an arbitrary integration constant. For the scale factor $a$
we can set
\begin{equation}
a\propto t^{2/{3\gamma_T}},
\end{equation}where the value of an integration constant has been fixed so that $\lim_{t\to 0}a=0$. 

 The solution will be inflationary if $\rho_T+3p_T<0$, and in terms of the tachyon field such condition is equivalent to 
\begin{equation}
\dot T^2<\frac{2}{3},
\end{equation}
which holds provided 
$\beta>{2}/{\sqrt{3}}$.

\subsection{Tracking solutions}
We move on now to the most interesting case, which is represented by the attractor solutions
with $(x,y)=(\sqrt{\gamma},\gamma/\alpha)$. They satisfy \begin{equation}\Omega_T=1-\Omega_{\gamma}=\frac{\gamma}{\alpha\sqrt{1-\gamma}},\end{equation} so that
the energy density of the tachyon and of the fluid scale exactly as the same power of the scale factor, namely $\rho_{\gamma}\propto\rho_{T}\propto a^{-2/3\gamma}.$
For that reason, according to the definition by Liddle and Scherrer \cite{liddle},  these solutions display a tracking behaviour. Now,  
as in the case of the tachyon dominated solutions the expansion factor obeys a power-law, 
$a\propto t^{2/3\gamma}$, and inflation will proceed if $\gamma<\rm{min}\{2/3,\alpha y_1\}$.

Nevertheless, apart from the considerations above, there exist restrictions on the  values of $\alpha$ (or $\beta$) that come from observations. Using standard nucleosynthesis and the observed abundances of primordial nuclides, the strong constraint that the fractional energy density of scalar matter cannot exceed $0.05$ at temperatures near $1$ MeV was set
\cite{bean}. If we restrict the discussion to tracking inflationary solutions ($\gamma<2/3$),
and set $\Omega_T <0.05$, we see  $\alpha>23.09$ is required. Such bound, though, can be evaded if the solution begins its history away from the tracking solution
and in a region where $\Omega_T\ll1$ (close to $O$), and only reaches $Q$ in the course of the evolution.

\section{Conclusions}
The evolution equations of a spatially flat FRW universe containing a barotropic fluid plus a
tachyon $T$ with an inverse-square potential $V(T)=\beta T^{-2}$ define a two-dimensional flow. The evolution of such models has been investigated by studying the orbits of that flow in the physical state, which is in this case a subset of the Euclidean plane. The Friedmann constraint, combined with a careful choice of coordinates, renders this subset compact. 

We have shown that the energy density of the tachyon dominates at late times for  $\gamma>\alpha(\sqrt{\alpha^2+4}-\alpha)/2$, where $\gamma$ is the barotropic index of the fluid and $\alpha=4/3\beta$. In constraint, for $\gamma<\alpha(\sqrt{\alpha^2+4}-\alpha)/2$, the barotropic fluid does not dominate completely and the contribution of tachyonic energy density to the total one is not negligible. Nucleosynthesis imposes, then, tight bounds on the admissible values of $\alpha$, but such restrictions can be relaxed if the locus of the initial solution is far from that of the tracking one
and in a region where $\Omega_T\ll1$ (close to $O$), and only reaches $Q$ in the course of the evolution.

Finally, a possible generalization of this work would be considering generalized tachyon cosmologies 
like those presented in \cite{Chimento}.

\section*{Acknowledgments}
JMA and RL are supported by the University of the Basque Country through research grant 
UPV00172.310-14456/2002. JMA also acknowledges support from the Spanish Ministry of Science and Technology through research grant  BFM2000-0018. RL is also is supported by the Basque Government through fellowship BFI01.412, the Spanish Ministry of Science and Technology
jointly with FEDER funds through research grant  BFM2001-0988.

%%%%%%%%%%%%%%%%%%%%%%%%%%%%%%%%%%%%%%%%%%%%%%
% Bibliography
%%%%%%%%%%%%%%%%%%%%%%%%%%%%%%%%%%%%%%%%%%%%%%

%%%%%%%%%%%%%%%%%%%%%%%%%%%%%%%%%%%%%%%%%%%%%%
% Figures
%%%%%%%%%%%%%%%%%%%%%%%%%%%%%%%%%%%%%%%%%%%%%%

%%%%%%%%%%%%%%%%%%%%%%%%%%%%%%%%%%%%%%%%%%%%%%
% Figure 1
%\newpage
%\begin{figure}
%\begin{center}
%\includegraphics[width=0.4\textwidth]{phase.eps}
%\end{center}
%\caption{Toroidal solenoid and point charge.\label{fig:phase}} 
%\end{figure}

%%%%%%%%%%%%%%%%%%%%%%%%%%%%%%%%%%%%%%%%%%%%%%
% Figure 2
%\newpage
%\begin{figure}
%\begin{center}
%\includegraphics[width=\textwidth]{g12a15.eps}
%\end{center}
%\caption{Phase-space for $\alpha=1.5$ and $\gamma=1.2$.\label{fig:g12a15}} 
%\end{figure}

%%%%%%%%%%%%%%%%%%%%%%%%%%%%%%%%%%%%%%%%%%%%%%
% Figure 3
%\newpage
%\begin{figure}
%\begin{center}
%\includegraphics[width=\textwidth]{g03a15.eps}
%\end{center}
%\caption{Phase-space for $\alpha=1.5$ and $\gamma=0.3$.\label{fig:g03a15}} 
%\end{figure}

\end{document}